\documentstyle[12pt]{article}

\newcommand{\be}{\begin{equation}}
\newcommand{\ee}{\end{equation}}
\newcommand{\bea}{\begin{eqnarray}}
\newcommand{\eea}{\end{eqnarray}}
\newcommand{\bean}{\begin{eqnarray*}}
\newcommand{\eean}{\end{eqnarray*}}

\def\lapproxeq{\lower .7ex\hbox{$\;\stackrel{\textstyle <}{\sim}\;$}}
\def\gapproxeq{\lower .7ex\hbox{$\;\stackrel{\textstyle >}{\sim}\;$}}

\def\asp{(\frac{\alpha_s}{\pi})}

\begin{document}
\begin{titlepage}
\vspace*{-1cm}
\begin{flushright}
RAL-93-040\\
June 1993
\end{flushright}
\vskip 1.cm
\begin{center}
{\Large\bf Consistent Analysis of the Spin Content of the Nucleon}

\vskip 1.cm
{\large F.E. Close} and {\large R.G. Roberts}
\vskip .2cm
{\it
Rutherford Appleton Laboratory,  \\
Chilton, Didcot  OX11 0QX, England
} \\
\vskip .4cm
\vskip 1cm
\end{center}

\begin{abstract}
The recent measurements of lepton nucleon scattering with polarised neutron
and deuteron targets are analysed together with the previous
polarised proton data in a mutually consistent way.
The detailed $x$-dependence of the polarisation asymmetry in the valence
region is shown to be in agreement with historical predictions based on
quark models.
The Bjorken
sum rule is shown to be confirmed at the $1\sigma$ level
and estimates of the spin content of the nucleon
$\Delta q$ are extracted. While the average value of $\Delta q$ from the
three experiments comes out to be $0.41\pm0.05$ (to be compared with the
naive quark model theoretical expectation of 0.58) this experimental average
value is more than one
standard deviation from the value obtained from any individual experiment.
This inconsistency
can be overcome by allowing arbitrary higher twist contributions but the
resulting precision is poor, $\Delta q = 0.38\pm 0.48$.
\end{abstract}
\vfill
\end{titlepage}
\newpage

\section*{Introduction}

The recent measurements of the polarised nucleon structure functions $g_1$ for
the deuteron\cite{SMC} and the neutron\cite{SLACE142} have re-kindled the
debate over the spin content of the nucleon which began with the measurement
of $g_1$ for the proton\cite{EMC} five years ago.  The value of $I_p=\int
gp_1(x){\rm d}x$ extracted in ref\cite{EMC} was consistent with a tiny
fraction
  of the
proton's spin being carried by the constituent quarks and this fuelled enormous
speculation over our understanding of the nucleon in the quark model
framework.  Reviews of the various interpretations of this result and of the
competing descriptions of the proton's spin structure can be found in ref
\cite{reviews}.

In order to draw conclusions from the three experiments it is important to
compare them consistently, in particular at the same $Q2$ and with the same
ancillary inputs (e.g. the unpolarised $F_1(x,Q2)$ used in constructing the
polarised structure function $g_1(x,Q2)$ from the measured asymmetry
$A_1(x)$).
It is the purpose of the present paper to do this. A central plank in our
analys
 is
will be the asymmetry $A_1(x)$ and we begin with a comment on this measured
quantity.

A significant feature of the data is that the $x$-dependences of the {\it
polari
 sation
asymmetry} in the valence region, $A_1(x>0.2)$, confirm the quark model
predictions\cite{VQM}\cite{KW}
for proton, neutron (see fig.1) and deuteron systems.
This suggests to us that there is an immediate message from these data:
\begin{center}
{\bf The polarisation of the {\it valence} quarks is canonical}
\end{center}
and this should be taken into
account in any attempt to interpret the data. The $A(x)$ has tended to be
ignored in the literature while
most of the attention, and associated controversy, has arisen from
the value of the {\it integrated
  structure function} $g_1(x)$ and its interpretation.
Much of our paper will address the
implications of the new data for this question.

One obvious intention of a simultaneous analysis of $g_1p,gn_1$ and $gd_1$
is to compare the experimental estimate of $I_{p-n}$ with Bjorken's
fundamental sum rule\cite{Bj}
\be
I_{p-n}(Q2) \equiv \int1_0  (g_1p(x,Q2)-g_1n(x,Q2))\;{\rm d}x =
\frac{1}{6}\frac{g_A}{g_V}\;\left [ 1 - \asp - \frac{43}{12}\asp2 \right ]
\label{bjsr}
\ee
where we assume $n_f = 3$.
  Since the three experiments carry out measurements at
different values of $Q2$ one must be careful to combine the $p, n, d$ results
a
 t a
common $Q2$ to test the Bjorken sum rule.  For this reason and for general
requirements of consistency we shall take only the measurements of the {\it
asymmetry} from refs\cite{SMC, SLACE142, EMC} and use the latest sets of
unpolarised structure functions and parton distributions to construct the
polarised structure functions through
\be
g_1(x) = A_1(x) F_1(x) = \frac{A_1(x)F_2(x)}{2x(1+R(x))}  \label{g1def}
\ee
We find that $gp_1$ (and $I_p$)
of ref\cite{EMC} is
increased as a result of the new information on $F_2(x)$ from ref\cite{NMC} at
low $x$.

Also, in extracting `experimental' estimates of the integrals $I_{p,n,d}$ we
are guided by theoretical estimates of the asymmetry $A_1$ at {\it large $x$}
to cover the unmeasured region ($x > 0.6$). Even where the asymmetry is
measured for $x > 0.3$ the experimental uncertainty tends to be large and
can dominate the error on the integral (particularly SMC $d$ data),
so we prefer instead to use the valence quark model
(VQM) estimates of $A_1$ in this region also.

Comparison of the $I_{p,n,d}$ with the Ellis-Jaffe\cite{EJ} predictions and
the extraction of the spin content $\Delta q$ of the nucleon
require knowledge of the F/D parameter and careful treatment of QCD
corrections.
We re-evaluate F/D in the light of recent $\beta$-decay measurements.  We
include the QCD corrections to the non-singlet and singlet contributions to the
integrals.  Indeed in the non-singlet case the corrections
are known to second order at least\cite{Larin} and significantly reduce the
magnitude predicted for the Bjorken sum rule at low $Q2$.

We find that analysing the data in this manner is consistent
with the Bjorken sum rule at the $1\sigma$ level.
We find $\Delta q = 0.41\pm0.05$ but the values from each of
$p$, $n$ and $d$ lie outside the uncertainty of this mean value.
Allowing for higher-twist
contributions of arbitrary strength to force a common value of $\Delta q$ from
$p,n$ and $d$ leads to $\Delta q = 0.38 \pm$ 0.48. The errors
on the higher twist terms themselves are thus large and, not surprisingly,
are consistent with the rather precise theoretical
estimates of ref\cite{BBK} used in the recent analysis of Ellis and
Karliner\cite{EK} which yields $\Delta q = 0.22\pm 0.10$.

\section*{Extraction of $g_1{p,n,d}$ and $I_{p,n,d}$ from data}

The EMC proton experiment is at $<Q2> \sim$ 11 GeV$2$,  the SMC deuteron
experiment is at $<Q2> \sim$ 5 GeV$2$ while the SLAC E142 experiment is at
$<Q2> \sim$ 2 GeV$2$.  To evaluate the structure functions at a common $Q2$
value of 5 GeV$2$ we take the measured values of the asymmetries
$A_1{p,n,d} (x)$ for each experiment and assume these values hold
at $Q2$ = 5
GeV$2$.
(There is excellent evidence for the $Q2$ independence of $Ap_1(x)$ from ref
\cite{EMC} over the range 0.5 - 50 GeV$2$; within the relatively large errors
o
 f
ref\cite{SMC} there is no evidence for any $Q2$ dependence of $Ad_1(x)$.
Furthermore, within the precision of the SLAC E142 experiment $A_1n(x)$
appears to also to be independent of $Q2$\cite{emlyn}).
To construct the $gi_1(x,Q2)$ at $Q2$ = 5
GeV$2$ we take parton distributions to compute $F_1{p,n} (x,Q2)$ which are
consistent with recent DIS data, in particular the  $F_2$ data of NMC
\cite{NMC} at small $x$.  We take the $D\prime_0$ or $D\prime_-$
distributions
  of MRS
\cite{MRS}, the latter even providing an excellent description of the new data
from HERA\cite{HERA}.  The reliable estimate of the gluon distribution in
these fits provides, in turn, an estimate for $R_{QCD}$ to insert in
eqn(\ref{g1def}).  We have checked that these results remain true when
the distributions of ref\cite{CTEQ} are used instead.

The use of up-to-date unpolarised structure functions changes the values of
$gp_1(x)$ based on the EMC measurements at $Q2 \sim$ 11 GeV$2$ by a
significant amount;  it is essentially the new information from
NMC\cite{NMC} measurements which increases the
values of $gp_1(x)$ and hence $I_p$ and $\Delta q$.
  Computing the $g_1{p,n,d}(x)$ at the same $Q2$ value now
allows us to take combinations of pairs.  In fig.2 we compare the values of
$xgd_1(x)$ computed from the $Ad_1$ of SMC with the combination
$\frac{1}{2}(xgp_1(x)+ xgn_1(x))$ computed from the EMC and SLAC
asymmetry measurements.  Point-by-point we see that the two estimates are
consistent with each other.

In order to compute the integrals $I_{p,n,d}$ at $Q2$ = 5 GeV
we must extrapolate at small $x$ and large $x$.  The small $x$
estimate of the integral is obtained by taking the smallest $x$ data point and
assuming the behaviour of $xg_1\sim x\alpha$.  Taking $\alpha = 0$ (as
expected from Regge behaviour) gives the central value of this extrapolation
and

the error on this is the value obtained if $\alpha$ = 0.5.  Given that
the HERA data on $F_2(x)$
are {\it larger} than naive Regge expectation, one may need to reevaluate the
$g_1(x \rightarrow 0)$ extrapolation: our error estimate
allows for some room in this direction.  It is crucial that future
experiments go to as small $x$ as possible in order to help settle this
question
 .
Fig.2 indicates that the estimate for
the integral $\int{0.03}_0 gd_1(x)dx$  from the $\frac{1}{2}(p+n)$
combination

is rather different from the direct $d$ data.

At large $x$ we have some theoretical guidance for the asymmetries
$A{p,n}_1(x)$ from valence quark models\cite{VQM}.  Indeed, independent of
the questions about the values of the $I_i$, the localised $x$-dependence in
the valence region provides rather dramatic confirmation of predictions made
 far in advance of data on $n,d$ even $p$. We regard this as an important clue
in interpreting the polarisation data and therefore draw attention to, and make
a brief comment on, this aspect of the data which has tended to receive less
attention than the integral.

As $x\rightarrow 1$
both $Ap_1, An_1$ were predicted to reach
unity\cite{VQM}\cite{FEC73}\cite{COU
 NT}
 but their values around $x$ = 0.5
are expected to be quite different\cite{VQM}\cite{KW}.
  The expectations from the VQM are consistent
with the measured values at $x$ = 0.35, 0.45.  We therefore use the VQM
estimate
 s
of $A_1{p,n}$ (with estimated uncertainties) to compute the large $x$ integral
(i.e. $x >$ 0.6).  In addition we notice that the final errors in $I_{p,n,d}$
te
 nd to be
dominated by the last two values of the $g_1$ at $x$ = 0.35, 0.45 where the
cros
 s-
sections are small.  With some caution, we choose to take the VQM values with
their smaller uncertainties at these two $x$-values.  Fig.2 also shows these
values and we see that they are completely in line with the relatively precise
values obtained from the $\frac{1}{2}(p+n)$ combination.

At $Q2$ = 5 we compute the integrals and as a result of the above procedures
obtain the following values:-
\newpage
\bea
I_p(Q2 = 5) &=& \;\;\;0.135 \pm 0.011\nonumber\\
I_n(Q2 = 5) &=& -0.028 \pm 0.006\nonumber\\
I_d(Q2 = 5) &=& \;\;\;0.041 \pm 0.016  \label{IEXP}
\eea

Note the value of $I_p$ is larger than the EMC quoted value (due to new NMC
measurements of $F_2$) and note the larger central value of $I_d$ compared to
that quoted by SMC - due to our model estimates at large $x$ (but $I_d$ =
0.041
is within the quoted SMC uncertainty of course).  From these values we can
estimate in three ways the value of the Bjorken sum rule (where $d = (p+n)/2$)
at $Q2 = 5 GeV2$:-
\bea
I_{p-n} = 0.163 \pm 0.013\nonumber\\
I_{2(d-n)} = 0.139 \pm 0.035\nonumber\\
I_{2(p-d)} = 0.187 \pm 0.040\label{IP-N}
\eea

\section*{Bjorken Sum Rule and Spin Content of the Nucleon}

We can write for the first moments:-
\bea
I_p &=& \;\;\;I_3 + I_8 + I_0\nonumber \\
I_n &=&- I_3 + I_8 + I_0\nonumber \\
I_d &=& \;\;\;\;\;\;\;\;\;\; I_8 + I_0\label{IPND}
\eea
where
\bea
 I_3 &=& \frac{1}{12} a_3 (1-\frac{\alpha_s}{\pi} - 3.58
(\frac{\alpha_s}{\pi})
 2)
\nonumber \\
 I_8 &=& \frac{1}{36} a_8 (1-\frac{\alpha_s}{\pi} - 3.58
(\frac{\alpha_s}{\pi})
 2)
\nonumber \\
 I_0 &=& \;\frac{1}{9}\; a_0 (1-\frac{\alpha_s}{3\pi})\label{I380}
\eea
where the next-to-leading order QCD corrections to the non-singlet quantities
have been evaluated in ref\cite{Larin}.  Note that the QCD corrections to the
singlet are smaller.
\footnote{This is due to the non-vanishing of the singlet anomalous dimension
$\gamma_{qq}{(1),S,1}$\cite{Kod}. The $O(\alpha_s)$ correction to the singlet
coefficient function has recently been carried out\cite{ZVN}.}
  In eqn(\ref{I380}) $a_3$ and $a_8$ are related to the F/D
values while $a_0$ is the spin fraction carried by quarks, i.e.
\bea
a_3 &\equiv& \;F+D\; \equiv \Delta u - \Delta d\nonumber\\
a_8 &\equiv& 3F-D \equiv \Delta u + \Delta d-2\Delta s\nonumber\\
a_0 &\equiv& \;\;\;\;\Delta q\;\;\;\; \equiv \Delta u + \Delta d+\Delta s
       \label{a380}
\eea
Thus we need to know F, D precisely to extract a reliable estimate for $\Delta
q
 $.

We have performed a fit of the current values of the $\beta-$decay constants
for $np$, $\Lambda p$, $\Xi \Lambda$ and $\Sigma n$\cite{fd1} and
the best fit, with $\chi2$ = 1.55 for one degree of freedom (F/D with F+D
constrained to equal 1.257) is
\be
\left. \begin{array}{lll}
F& = & 0.459\pm 0.008\\
 D &=& 0.798\mp 0.008\end{array}
\right\} F/D
= 0.575 \pm 0.016
\label{FD}
\ee
This is 1$\sigma$ larger than the value used in a previous analysis of ours
\cite{fd2} principally due to
improved $\Lambda p$ and $\Sigma n$ data.  We shall use these values in what
follows; however there are two caveats.  First there is a systematic error, not
included, whch arises from the phase space or form factor corrections in the
$\Delta S$ = 1 examples\cite{fd3}.  The second is potentially more serious.

The quoted figures assume that in the hadronic axial current
\be
A_\mu = g_A\gamma_\mu \gamma_5 - g_2 \frac{i\sigma_{\mu\nu}q\nu
\gamma_5}{m_i+m_j}
\ee
one has $g_2$ = 0.  While this is assured in the limit where $m_i=m_j$ (such as
$n\rightarrow p$) it is not necessarily so for $\Delta S=1$.  Indeed, in quark
models one expects that $g_2=0 (\frac{m_i-m_j}{m_{ij}}$) with $m_{ij}\equiv
\frac{1}{2}(m_i+m_j)$\cite{fd4}.

Hsueh et al.\cite{fd5} made a fit allowing for $g_2\neq 0$
and found a significant change in their inferred value for $g_A$. A best fit
incorporating this value raises a tantalising possibility that the $(g_A/g_V$)
t
 hroughout the octet are
given by the naive quark model, all values being renormalised by 25\% (such
that the net spin is 0.75 rather than 1).
Such an eventuality would correspond to the realisation of the
effective quark model result
\be
F= 1/2, \; D = 3/4 \quad ; \quad F/D = 2/3
\ee

This is discussed elsewhere\cite{BIBLE}. These
 additional theoretical uncertainties merit further study: for the purpose
of comparing most directly with the literature we shall adopt the $g_2=0$ best
fit, eq(\ref{FD}).

  Since the error
on F+D is so small, the uncertainty on the Bjorken sum rule prediction comes
only from the $\alpha_s$ uncertainty:we take
 $\alpha_s$ = 0.28 $\pm$ 0.02 at $Q2$ = 5 GeV$2$
 which gives $I_{p-n}$ = 0.183
$\rightarrow$ 0.187.  We note that actually the $0(\alpha_s3)$ correction
estimate in ref\cite{Larin} is again, like the coeffs of $\alpha_s$ and
$\alpha_s2$, negative.  In fig.3(a) we compare our estimates from
eqn(\ref{IP-N})
with this value and we see no serious discrepancy.  Remember that our
procedure at large $x$ produces uncertainties which are smaller than the true
experimental errors - using the latter would further reduce
any possible discrepancy in
fig.3(a).

With the above values of F/D and $\alpha_s$, the Ellis-Jaffe\cite{EJ}
prediction
 s
(i.e. $\Delta s =0$) of the integrals $I_{p,n,d}$ are
\newpage
\bea
I_p &=&\;\;\; 0.172 \pm 0.009\nonumber\\
I_n &=& -0.018 \pm 0.009\nonumber\\
I_d &=&\;\;\; 0.077 \pm 0.009\label{Jaffe}
\eea
and these are compared with the values from eqn(\ref{IEXP}) in fig.3(b).
Despite the
increased estimate of $I_p$ from the EMC data we see that only $I_n$ is
consistent with the assumption of $\Delta s=0$.
Extracting the values of $\Delta q$ from
eqns (\ref{IEXP},\ref{IPND},\ref{I380},\ref{a380}) gives
\bea
I_p  \Rightarrow \Delta q = 0.21 \pm 0.11 \qquad (\Delta s = - 0.12 \pm
0.04)\nonumber \\
I_n  \Rightarrow \Delta q = 0.49 \pm 0.06 \qquad (\Delta s = - 0.03 \pm
0.02)\nonumber \\
I_d  \Rightarrow \Delta q = 0.24 \pm 0.15 \qquad (\Delta s = - 0.11 \pm
0.05)
\eea
to be compared with the theoretical expectation $\Delta q = 3F-D = 0.58$;
these are shown in fig.3(c).  The mean value of the spin content
from the three
determinations is
\be
<I_{p,n,d} > \;\;\; \Rightarrow \;\;\; <\Delta q> \;\;\;= \;\;\;0.41 \pm 0.05
\ee
the value being driven largely by
the relatively small errors on the neutron estimate.

Although these determinations agree at $1\sigma - 2\sigma$,
it is nonetheless somewhat unsatisfactory that the three determinations fail
to give a mutually consistent
value of the nucleon spin content. We note that the $p,d$, which are at
moderate
 ly
high values of $Q2$ agree while it is the neutron data at low $Q2$ that
appear to be out of line. This calls into
question the assumption made in the determination of the $g_1$ from the data
on $A_1$ and suggests that there may  indeed be significant $Q2$ dependence of
 $An_1(x)$
between $Q2$ = 2 and 5 GeV$2$ which we have neglected.  We know that
leading twist alone is insufficient to explain the unpolarised structure
functio
 ns
for $Q2$ below $\sim$ 4 GeV$2$ and therefore it seems sensible to allow for
some arbitrary higher twist
contributions to the first moments of $g_1$. Hence we write
\bea
I_p  &=& I_3 + I_8 + I_0 + a_p/Q2  \nonumber \\
I_n  &=& -I_3 + I_8 + I_0 + a_n/Q2  \nonumber \\
I_d  &=&  I_8 + I_0 + (a_p+a_n) /2Q2  \label{IHT}
\eea

We now evaluate each integral at the relevant $Q2$, i.e. $I_p$
at $Q2$ = 10.7 GeV$2$, $I_d$
at $Q2$ = 4.6  GeV$2$ and $I_n$ at $Q2$ = 2 GeV$2$. \footnote{While we
are now stressing possible $Q2$ dependence we should be concerned that even
within the EMC and SMC experiments there is a different $Q2$ range for each
$x$-value.  If this is true also for the SLAC experiment then our higher twist
analysis could be affected.}  We still use the MRS\cite{MRS} distributions for
$F_1(x,Q2)$ for determining $g_1{p,d}$ but, as these are not valid below 5
GeV$2$, for the determination of $g_1n(x, Q2=2$) we use both the NMC
\cite{NMC} and SLAC\cite{Whitlow} data on $F_2$ and $R$.  As a result, we
now get
\bea
I_p(Q2=10.7) &=& \;\;\; 0.134 \pm 0.012 \nonumber\\
I_n(Q2=\;\;2.0) &=& -0.023 \pm 0.005 \nonumber\\
I_d(Q2=\;4.6) &=& \;\;\; 0.041 \pm 0.016   \label{NEWIPN}
\eea
Taking $\alpha_s$ = 0.26 at $Q2$ = 10.7,
$\alpha_s$ = 0.36 at $Q2$ = 2 GeV$2$ we can extract the values of $a_p$ and
$a_n$ from eqn (\ref{IHT}) by insisting on a {\it common} value of $\Delta q$
from all three equations, using the same values of F and D as before.

As a result we obtain
\be
\Delta q = 0.38 \pm 0.48
\ee
with
\bea
a_p &=& -0.161 \pm 0.530\nonumber\\
a_n &=&\;\;\; 0.030 \pm 0.104    \label{DQHT}
\eea
We see that the errors on $I_{p,n,d}$ are such that the higher twist
contributions and $\Delta q$ cannot be pinned down with any precision.
The above values in eqn(\ref{DQHT}) easily encompass the QCD sum rule estimates
of Balitsky et al\cite{BBK} used in the analysis of Ellis and
Karliner\cite{EK},

\bea
a_p &=& -0.005 \pm 0.040\nonumber\\
a_n &=&\;\;\; 0.039 \pm 0.040    \label{EKHT}
\eea
The analysis of ref\cite{EK} concluded that $\Delta q = 0.22\pm 0.10$
and we can see that our analysis indicates the sensitivity of this result
to the magnitude of the higher twist terms. Also we see that the higher
twist contribution can make a substantial reduction to the magnitude of
the Bjorken sum rule. We note also that $O(1/Q2)$ terms occur naturally when
the Bjorken and Drell-Hearn-Gerasimov\cite{DHG} sum rules are derived in
explicit quark models\cite{Li}. Indeed their magnitudes are consistent
with the general bounds of eqn(\ref{DQHT}).

An interesting consequence of the values of $a_p$ and $a_n$ in eqn(\ref{EKHT})
is that it is primarily the neutron which would be expected to be most affected
at low $Q2$ leading to possible
dramatic effects for $gn_1(x,Q2)$.  With the above value for
$a_n$, $I_n$ would be  expected to change by
around 50\% between $Q2$ of 2 and 5 GeV$2$
and we can speculate how $gn_1(x,Q2)$ itself would alter to bring this
dramati
 c
increase in the size of $I_n$.  At present the only guide we have is a
compariso
 n
of [$xg_1d(x,Q2 =5) - xgp_1(x,Q2=5)]$  extracted from SMC and EMC.
There is a hint from this comparison that the increase
would occur at very low $x$,
especially if we keep faith with the VQM prediction for $x>$ 0.3.  However
since the SLAC data stop at $x$ = 0.03 it is also possible that $gn_1$ may be
much larger in magnitude for $x<$ 0.03, even at $Q2$ = 2 GeV$2$, causing
$I_n$ at $Q2$ = 2 GeV$2$ to be larger than supposed.
In that case any higher twist analysis would have to be severely modified.

Our analysis shows the importance of continued experimentation  in this
area.
The situation should become
clearer when SMC (who reach the smallest $x$ values of all three experiments)
have accumulated their full data sample and when SLAC are able to continue
their experiment for a polarised proton target.  We have stressed the
importance

of comparing $g{p,n,d}_1(x,Q2)$ and their integrals $I_{p,n,d}$ at the same
value of $Q2$ in order to pin down the spin content of the nucleon.  The
exercise of including higher twists in the analysis of present data indicates
the extreme sensitivity of the spin content of the nucleon to their
magnitude.

In conclusion we reiterate that the successful quark model predictions of the
$A(x>0.1)$\cite{VQM,KW} imply that valence quarks are polarised canonically
and that there is no need to rewrite the textbooks in light of these data.
Immediate questions to be answered include whether $An(x\rightarrow 1) >0$
\cite{VQM}\cite{COUNT};
with this exception the behaviour of $A(x)$ in the valence region seems
establis
 hed
and most effort is needed in the $x \rightarrow 0$ region.
In addition to the questions
advertised above, we urge test of whether $g_1d(x \rightarrow 0) <0$ as this
may enable a ``direct" measure of $\Delta q$\cite{fd2,BIBLE}. Finally, if
the valence quarks are indeed polarised canonically then it becomes important
to
make direct measure of the sea polarisarion. Ref.\cite{FM} has argued that
semi-inclusive production of fast $K-$ in polarised leptoproduction may
enable the polarisation of $s$ and/or $\bar{u}$ to be probed.

\section*{Acknowledgment}

We thank Graham Ross for discussions and suggestions.

\bigskip

\noindent{\Large\bf Figure Captions}
\begin{itemize}

\item[{[1]}] Polarisation asymmetries $Ap_1(x,Q2)$ and $An_1(x,Q2)$
from the EMC\cite{EMC} and SLAC E142\cite{SLACE142} experiments compared
with the predictions of valence quark model\cite{VQM}. The
curves\cite{VQM} correspond to $Ap_1 = \frac{19-16R}{15}\xi$,
$An_1 = \frac{2-3R}{5R}\xi$, with $R = \frac{F_1n}{F_1p}$ and
$\xi =1$ (solid), $\xi = 0.75$ (dashed). See ref\cite{BIBLE} for further
details.

\item[{[2]}] $xg_1d(x,Q2=5)$ extracted from the SMC data on $Ad_1(x)$
and $g_1{\frac{1}{2}(p+n)} (x,Q2=5)$ from the corresponding data on
$A{p,n}_1(x)$ from EMC and SLACE142.  Also included are the values at large
$x$ expected from the valence quark model.

\item[{[3(a)]}] Values of Bjorken sum rule $I_{p-n}$ extracted from the values
o
 f
$I_{p,n,d}$ at $Q2$ = 5 GeV$2$.  The shaded region is the theoretical
prediction.\\

\item[{[3(b)]}] $I_{p,n,d}$ extracted at $Q2$ = 5 GeV$2$ compared to the
expec
 tation of the
Ellis-Jaffe sum rules, $\Delta s=0$.\\

\item[{[3(c)]}] Values of $\Delta s$ and $\Delta q = \Delta u + \Delta d +
\Delta s$ extracted
from the estimates of  $I_{p,n,d}$ at at $Q2$ = 5 GeV$2$.\\

\end{itemize}

\end{document}